\documentclass[preprint,showpacs,prl]{revtex4}
\usepackage{graphicx}
\usepackage{amsmath}
\begin{document}
\title{Edge-state Fabry-Perot interferometer as a high sensitivity charge
detector}
\author{P. K. Pathak and Kicheon Kang}
\affiliation{Department of Physics, Chonnam National University,
Gwangju 500-757, Republic of Korea}
\date{\today}
\begin{abstract}
We present a scheme for high sensitivity charge detection in the
integer quantum Hall regime using two point contacts in a series.
The setup is an electronic analog of an optical Fabry-Perot
interferometer. We show that for small transmission through the
point contacts the sensitivity of the interferometer is very high
due to multiple reflections at the point contacts. The sensitivity
can be further enhanced twice by using electrons in spin entangled
state. We show that for point contacts having different reflection
probabilities, the interferometer can be tuned for the quantum
limited measurement.
\end{abstract}
\maketitle

Measurement of the charge-state of a mesoscopic system has generated
lot of interest in recent years \cite{meas,buks98,chang08}, mainly
due to the applications of charge qubits in solid-state realization
of quantum information processing~\cite{qinf}. Mesoscopic devices
such as quantum point contact (QPC)~\cite{qpcd} and single electron
transistor (SET) \cite{setd} have been widely used as the charge
detectors. These detectors do not perform instantaneous measurement,
but the measurement is performed as a sequence of continuous weak
measurements \cite{weak}. The merits of these detectors can be
understood from the two points of view : (1) efficiency and (2)
sensitivity. The former is related to the back-action noise produced
by the detector and the latter is related to the precision. The
quantum mechanical complementarity establishes a trade-off between
acquisition of information about the state of the system and the
back-action dephasing. A detector is called 100\% efficient
(quantum-limited) if the dephasing occurred in the measured system
is only due to the acquisition of information by the detector. 
Performing more
sensitive measurements have often led to reveal new physics
\cite{loyd}. A high sensitivity charge detector working in the
quantum limit can have wider applications in quantum metrology
\cite{metr}. The improvements in measurements can be accomplished
either through new designs of measurement devices or by developing
methods that rely on properties like correlations \cite{butt2} and
entanglement \cite{yurke,ynlee}.

\begin{figure}
\includegraphics[width=2.5in]{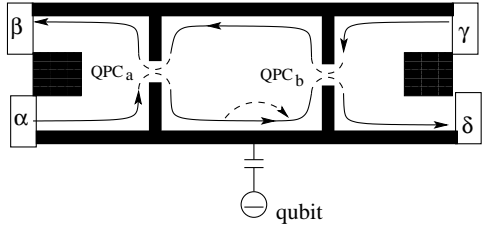}
\caption{Schematic arrangement for measurement of charge qubit.
Two spatially separated point contacts form the Fabry-Perot
interferometer. The qubit is capacitively attached in one arm of the
interferometer.} \label{fig1}
\end{figure}

In this Letter, we present an interferometry model of a high
sensitivity charge detector in the integer quantum Hall regime~\cite{mach}. 
For fractional quantum Hall states, a similar arrangement has been proposed 
for measuring fractional charge and non-Abelian statistics \cite{add}.
Our model is an electronic analog of Fabry-Perot interferometer
\cite{fabp}. We show that the charge sensitivity of our model is
higher than a two-path interferometer due to multiple reflections of
electrons at QPCs. We report the possibility of tuning the
interferometer for quantum limited measurement for $R_a<R_b$, where
$R_a$ ($R_b$) is reflection probability of quantum point contact
QPC$_a$ (QPC$_b$) (cf.Fig.~\ref{fig1}). 
We note that,
two-path interferometer with edge channel (Mach-Zehnder interferometer)
has been realized
\cite{mach}, 
Further, the possibility of
quantum limited detection
of charge using Mach-Zehnder interferometer has also been 
proposed~\cite{qlimit}.

In Fig.~\ref{fig1}, we show a schematic setup, constructed using
electrical gates on a Hall bar, for measurement of charge. 
Our detector consists of two
QPCs, QPC$_a$ and QPC$_b$, arranged in a series. The input electrons
are injected from the source terminals $\alpha$ and $\gamma$. The
outgoing electrons are collected at the drain terminals $\beta$ and
$\delta$. In the quantum Hall regime, QPCs act as the beam splitters
for the incoming electrons. The point contact QPC$_a$ splits the
incoming edge-state current from source $\alpha$ into two parts with one
reflected back to the drain $\beta$ and the other transmitted to the
second point contact QPC$_b$. The edge-state beam on reaching at
QPC$_b$ is further split into two parts, one transmitted to the
drain $\delta$ and other part reached at QPC$_a$, where it is again
partially transmitted to drain $\beta$ and partially reflected back
to QPC$_b$ and so on. 
Thus our detector is analogous to optical
Fabry-Perot interferometry. A charge-qubit is capacitively attached
to the lower arm of the interferometer between the two QPCs. The
qubit, having two charge states $|0\rangle$ and $|1\rangle$, could
be a double-quantum-dot or a two path interferometer. There is no
electron transfer from the qubit to the interferometer. Due to Coulomb
interaction the charge on the qubit deflects edge-state in the lower
arm without changing transmission through QPCs, which modifies the
phase of the edge-state-current via the Aharonov-Bohm effect.

The information of the measured state of the qubit is reflected in
the electrons collected at drain reservoirs. We follow scattering
matrix analysis for input-output probability amplitudes. The scattering
matrix in terms of Fermi operators at $m$-th terminal
$c_m,~m=\alpha,~\beta,~\gamma,~\delta$ is written as follows:
\begin{equation}
\left(\begin{array}{c}c_{\beta}\\c_{\delta}\end{array}\right)=\left[\begin{array}{cc}\bar{r_i}&\bar{t_i}^{\prime}
\\\bar{t_i}&\bar{r_i}^{\prime}\end{array}\right]\left(\begin{array}{c}c_{\alpha}\\c_{\gamma}\end{array}\right)
\label{smatrix} ,
\end{equation}
\begin{eqnarray}
\label{sc1}
&&\bar{r_i}=r_a+\frac{t_at^{\prime}_ar_be^{i(\phi+\theta_i)}}{1-r^{\prime}_ar_be^{i(\phi+\theta_i)}},
\bar{t_i}=\frac{t_at_be^{i\theta_i}}{1-r^{\prime}_ar_be^{i(\phi+\theta_i)}},\\
&&\bar{r_i}^{\prime}=r^{\prime}_b+\frac{t_bt^{\prime}_br^{\prime}_ae^{i(\phi+\theta_i)}}{1-r^{\prime}_ar_be^{i(\phi+\theta_i)}},
\bar{t_i}^{\prime}=\frac{t^{\prime}_at^{\prime}_be^{i\phi}}{1-r^{\prime}_ar_be^{i(\phi+\theta_i)}},
\label{sc2}
\end{eqnarray}
where $\phi$ is the Aharonov-Bohm phase acquired by the electron along
one complete loop between QPCs and $\theta_i$ is the phase produced
by the qubit. The phase $\theta_i$ has two values corresponding to
different charge states of the qubit $|i\rangle,~i=0,1$.
Typical value of the phase difference $\Delta\theta=\theta_1-\theta_0$
generated by the Coulomb interaction is about 
$\Delta\theta=0.03$~\cite{chang08}.
Effectively, charge state of the qubit modifies the amplitude as
well as the phase of the transmission through the detector. All other
phases in scattering are included in the transmission amplitudes
$t_n$ ($t^{\prime}_n$) from the left (right) and the reflection
amplitudes $r_n$ ($r^{\prime}_n$) on the left (right) for QPC$_n$,
$n=a,b$.

First, we consider electrons are injected only from the source
terminal $\alpha$ and collected at the drain terminal $\delta$. The
transmission probability $\bar{T}_i$ ($=|\bar{t}_i|^2$) of the interferometer is given by
\begin{equation}
\bar{T}_i(\Phi_i)=\frac{T_aT_b}{1+R_aR_b-2\sqrt{R_aR_b}\cos{\Phi_i}},
\label{T}
\end{equation}
where $\Phi_i=\theta_i+\phi$+arg($r^{\prime}_ar_b$) and
$T_n=|t_n|^2=1-R_n$. Sensitivity of the transmission probability $T$
to variation in phase $\Phi_i$ makes it possible to measure the
charge state of the qubit. The transmission probability has
Lorentzian-like resonances when $\Phi_i$ is multiples of $2\pi$. The
half width at half maximum of the resonance is
$\Gamma_w\approx(1-\sqrt{R_aR_b})/(R_aR_b)^{1/4}$. The resonances
are narrower for larger values of $R_a$ and $R_b$, which provides
larger change in current for small variations in phase $\Phi_i$. The
phase sensitivity of the interferometer is determined by the phase
fluctuations due to intrinsic shot noise. In the linear regime, the
average source-drain current is $\langle
I_i\rangle=(e^2V/h)\bar{T}_i$ and the shot noise is given by
$S_i=(2e^3V/h)\bar{T}_i(1-\bar{T}_i)$, where $V$ is source-drain
voltage. For time interval $t$, the average number of electrons 
transmitted 
is $\langle N_i\rangle=\langle I_i\rangle t/e$ and the fluctuation of 
number of electron is $\langle(\Delta N_i)^2\rangle=S_it/(2e^2)$. 
Therefore, the rms phase fluctuation
\cite{yurke} for the interferometer is given by
\begin{equation}
\Delta\Phi_i \equiv 
\frac{\sqrt{\langle(\Delta N_i)^2\rangle}}{|\partial\langle N_i\rangle/\partial\Phi_i|}
=\sqrt{\frac{h}{eVt}}\frac{\sqrt{\bar{T}_i(1-\bar{T}_i)}}{|\partial \bar{T}_i/\partial\Phi_i|}.
\label{sen}
\end{equation}
From Eq.~(\ref{T}) and (\ref{sen}) one can calculate the sensitivity
of Fabry-Perot interferometer. We compare the sensitivity of
Fabry-Perot interferometer with a two-path (Mach-Zehnder)
interferometer for which transmission probability is cosine function
of the form
$\bar{T}_i(\Phi_i)=R_aR_b+T_aT_b+2\sqrt{R_aR_bT_aT_b}\cos{\Phi_i}$
\cite{mach}. Near the resonance, for $R_a\approx R_b$, the ratio of
$\Delta\Phi_i$ for Fabry-Perot interferometer to Mach-Zehnder
interferometer is approximately $T_a^{3/2}$. Clearly, Fabry-Perot
interferometer can be used as a very high precision charge detector
for smalltransmission probabilities $T_a$,~$T_b$.

In real devices, this high precision would be limited by the finite
source-drain bias voltage, because the phase $\Phi_i$ acquires an additional energy
dependent fluctuating part~\cite{chung}. Considering drift velocity
$v_d$ as constant along the edges, we can write energy dependence of
phase as
$\Phi_i(\epsilon)=\Phi_i(E_F)+\epsilon/E_c,~E_c=\hbar v_d/L$,
where $L$ is the length of one complete loop between the QPCs,
 $E_F$ is Fermi energy and $\epsilon$ is small energy difference
for electrons from Fermi level. 
The averaging of the energy
dependent fluctuations gives average transmission probability and
average shot noise, respectively, as
$\langle\bar{T}_i\rangle =
(eV)^{-1}\int_{-eV/2}^{eV/2}\bar{T}_i(\Phi_i(\epsilon))d\epsilon$,
$\langle{S_i}\rangle = {2e^3}/{h}\int_{-eV/2}^{eV/2}
\bar{T}_i(\Phi_i(\epsilon))(1-\bar{T}_i(\Phi_i(\epsilon)))d\epsilon$.
At small bias
($eV/E_c\ll\Gamma_w$), we find that $\Delta\Phi_i$ is changed by the
factor
$\left[1+(eV/E_c)^2(\Gamma_w^2-\Phi_i^2)/2(\Gamma_w^2+\Phi_i^2)^2\right]$
(for $-\pi<\Phi_i<\pi$).

In order to understand the measurement process and the back action
of the detector, we consider evolution of the state of the combined
system of detector and qubit. When an electron is
injected from source $\alpha$ and the initial state of the qubit is
$a_0|0\rangle+a_1|1\rangle$, the state of the combined
qubit-detector system evolves as
\begin{equation}
|\psi\rangle = (a_0|0\rangle+a_1|1\rangle)c^{\dag}_\alpha
|F\rangle\rightarrow a_0|0\rangle|\xi_0\rangle+a_1|1\rangle|\xi_1\rangle,
\end{equation}
where $|F\rangle$ denotes Fermi sea of all the electrodes  and
$|\xi_i\rangle=(\bar{r}_ic^{\dag}_\beta+\bar{t}_ic^{\dag}_\delta)|F\rangle$ for
$i=0,1$ are detector states. The final state of the qubit is given
by the reduced density matrix
$\rho=Tr_{det}|\psi\rangle\langle\psi|$, obtained after tracing over
the detector states. The dephasing of qubit can be expressed in
terms of off-diagonal elements of density matrix $\rho$ as
$|\rho_{01}(t)| = |\rho_{01}(0)|\exp{(-\Gamma_d t)}$,
where $\Gamma_d$, detector back action induced dephasing rate, is
given by \cite{buks98,qlimit}
$\Gamma_d = -h^{-1}\int d\epsilon 
\log{|\bar{r}_0\bar{r}_1^*+\bar{t}_0\bar{t}_1^*|}$. 
In the linear regime, for weak measurement
($|\bar{r}_0\bar{r}_1^*+\bar{t}_0\bar{t}_1^*|\sim1$), the dephasing
rate $\Gamma_d$ can be expanded in terms of the change in the
transmission probability, $\Delta T = |\bar{t}_{0}|^{2} -
|\bar{t}_{1}|^{2}$, and the change in the relative scattering phase
$\Delta\zeta=arg(\bar{t}_{1}/\bar{r}_{1})-arg(\bar{t}_{0}/\bar{r}_{0})$
as follows,
\begin{subequations}
\begin{eqnarray}
 \Gamma_{d} &=& \Gamma_{T} + \Gamma_{\zeta},  \\
 \Gamma_{T} &=& \frac{eV}{8h}\frac{(\Delta  T)^{2}}{T(1-T)}, \;
 \Gamma_{\zeta} = \frac{eV}{2h} T(1-T)(\Delta\zeta)^{2},
\end{eqnarray}
\label{apdeph}
\end{subequations}
where $T=( |\bar{t}_1|^2 + |\bar{t}_0|^2 )/2$. The information of
the state of qubit is reflected in the change of source-drain current.
Therefore only the information of the qubit in the part of dephasing
related to the change in current $\Gamma_T$ is utilized by the detector.
One can find that the measurement rate of the detector
$\Gamma_m$ is equal to $\Gamma_T$. However, the information
lost in the part of dephasing $\Gamma_\zeta$ goes undetected. For
a quantum limited detector it is necessary that the unutilized
information in phases should be eliminated, i.e. $\Delta\zeta=0$. In
a single QPC detector that obeys mirror reflection symmetry and time
reflection symmetry the relative phase between transmission and
reflection amplitude remains constant and change in relative phase
$\Delta\zeta=0$~\cite{weak,butt1,othr}.

From Eqs.~(\ref{sc1}) and (\ref{sc2}) change in
relative phases between transmission and reflection amplitude for Fabry-Perot
interferometer is given by
\begin{equation}
\Delta\zeta=arg\left\{e^{i\Delta\theta}\frac{\sqrt{R_a}-\sqrt{R_b}
e^{i\Phi_0}}{\sqrt{R_a}-\sqrt{R_b}e^{i(\Phi_0+\Delta\theta)}}\right\}.
\label{phi}
\end{equation}
For $R_a=R_b$, from Eq.~(\ref{phi}), we get
$\Delta\zeta=\Delta\theta/2+\pi$, for
$0>\Phi_0>-\Delta\theta/2$, and $\Delta\zeta=\Delta\theta/2$, otherwise.
\begin{figure}
\includegraphics[width=3.1in]{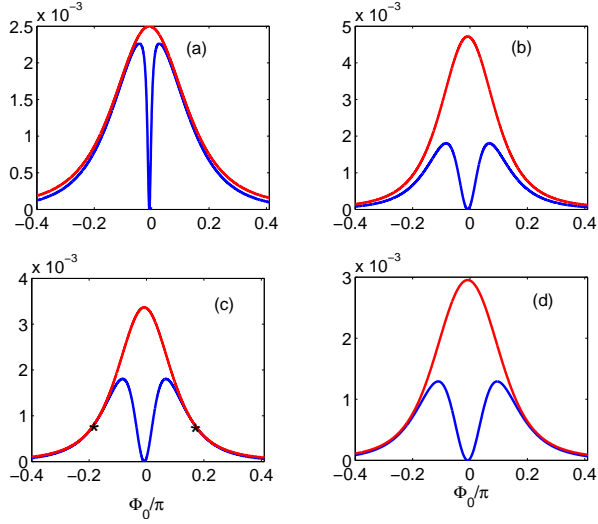}
\caption{(Color online) The renormalized measurement rate $\Gamma_m/\Gamma_0$ (blue line) and dephasing rate
$\Gamma_d/\Gamma_0$ (red line) for $\Gamma_0=eV/h$, $\Delta\theta=0.05$, and
(a) for symmetric interferometer ($R_a=R_b$=0.5), (b) for $R_a>R_b$ i
($R_a=0.7, R_b=0.5$), (c) for $R_a<R_b$ ($R_a=0.5, R_b=0.7$). 
The interferometer operates in
quantum limit for $\Phi_0=\pm\cos^{-1}\sqrt{R_a/R_b}$, shown as
black asterisks. (d) Same as (c) for small finite bias ($eV/E_c=0.5$).  
Note that $\Gamma_w\approx 0.53$ for $R_a=0.5$ and $R_b=0.7$.}
\label{fig2}
\end{figure}

In the case when both QPCs in Fabry-Perot interferometer have same
reflection probabilities $(R_a=R_b)$, $\Delta\zeta$ always remains
nonzero. Therefore there is always some information loss in the
phases which goes undetected and detector cannot perform quantum
limited measurement. Note that this behavior is different from the
detection with resonant transmission at zero magnetic
field~\cite{phas}, where the quantum-limited detection is possible only
for symmetric double QPCs. In Fig.~\ref{fig2}(a) we show measurement rate
$\Gamma_m$ and dephasing rate $\Gamma_d$ calculated from
Eq.~(\ref{apdeph}) for $R_a=R_b$. We find that dephasing rate of the
qubit is always higher than the measurement rate. In this case some
information is always lost in scattering phases, which means quantum
limited measurement is not possible. For higher values of $R_a$ and
$R_b$, detector has higher sensitivity and the measurement is nearly
quantum limited except at resonance. At resonance relative
scattering phase $\Delta\zeta$ faces an abrupt change by $\pi$ 
which results maximum loss of information. Further
because of the sensitivity of the detector is minimum at resonance,
the measurement rate faces dip. For smaller values of $R_a$ and
$R_b$ sensitivity of detector is smaller and more information is
lost in scattering phases. From Eq.~(\ref{phi}), change in relative
scattering phases for $R_a\neq R_b$ is given by
\begin{equation}
\Delta\zeta=\frac{\Delta\theta}{2}+\tan^{-1}\left[\frac{(R_a-R_b)
\sin(\frac{\Delta\theta}{2})}{(R_a+R_b)\cos(\frac{\Delta\theta}{2})-2\sqrt{R_aR_b}
\cos(\frac{\Phi_0+\Delta\theta}{2})}\right]. \label{nequal}
\end{equation}
In this case, we find the condition for quantum limited measurement
$\Delta\zeta=0$ simplifies to
${R_a}/{R_b}={\cos^2(\Phi_0+\Delta\theta/2)}/{\cos^2(\Delta\theta/2)}.
$
For small value of $\Delta\theta$, 
${\cos^2(\Phi_0+\Delta\theta/2)}/{\cos^2(\Delta\theta/2)}$ 
is always less than unity except at resonance where
quantum limited measurement is not possible. 
This clearly shows that in Fabry-Perot interferometer
quantum limited measurement can only be possible if $R_a<R_b$,
and the value of $\Phi$ for quantum limited
measurement is given by
$\Phi_0\approx\pm\cos^{-1}\sqrt{R_a/R_b}$. 
In Fig.~\ref{fig2}(b)-(d), we show dephasing rate and measurement
rate of qubit for Fabry-Perot interferometer having QPCs with
different reflection probabilities $(R_a\neq R_b)$. For $R_a>R_b$,
shown in Fig.~\ref{fig2}(b), dephasing rate is always larger
than the measurement rate. This shows that the detector has poor
efficiency for such construction. On the other hand, in
Fig.~\ref{fig2}(c) for $R_a<R_b$, there exist two points where the measurement
rate is equal to the dephasing rate at $\Phi_0\simeq\pm\cos^{-1}\sqrt{R_a/R_b}$.
These points are symmetrically placed on both sides of resonance.
For finite bias we average over the energy of the injected electrons.
We find that at small bias $eV/E_c=0.5\lesssim\Gamma_w$ (see
Fig.~\ref{fig2}(d)), our results are not modified much. 
The measurement rate is reduced very much at large biasing,
$eV/E_c\gg\Gamma_w$, and the quantum limited operation of the detector
is not possible. Similarly, we also found that (not shown here) thermal 
broadening at high temperature ($kT/E_c\gg\Gamma_w$) reduces the 
sensitivity and the efficiency. 

If we also include effect of environment on the qubit, the coupling to the environment relaxes the state of the qubit to its lower energy state. 
The condition when environment can produce dephasing and the measurement of relaxation rate has been discussed in detail in Ref.~\cite{relax}. Coupling of the qubit with environment can reduce the efficiency of the detector only when environment also produces dephasing.

Our findings are unique because of the following facts. 
For a single QPC as a quantum
limited charge detector, satisfaction of time reversal symmetry
and mirror-reflection symmetry is essential~\cite{weak,butt1,othr}. 
Technically construction of such
QPC may not be trivial, and the information loss is usually large for
generic QPC. The dephasing rate is reported about $30$ times larger than 
the measurement rate ~\cite{kalish04,kang05,chang08}. Here we report that 
in Fabry-Perot
interferometer quantum limited
measurement is possible only if the first QPC has smaller reflection
than the second QPC, ie $R_a<R_b$. Further, this Fabry-Perot construction 
provides much higher precision than a two-path (Mach-Zehnder) interferometer 
does.

Next, we briefly discuss improvement in sensitivity using quantum
entanglement. For our purpose we consider spin entangled singlet
pairs injected through identically biased input terminals $\alpha$
and $\gamma$. The state of injected electrons can be expressed as
$|\psi_{in}\rangle= \frac{1}{\sqrt{2}}
  \left(c_{\alpha\uparrow}c_{\gamma\downarrow}
   -c_{\alpha\downarrow}c_{\gamma\uparrow}\right)|F\rangle,
$
where $\uparrow$ and $\downarrow$ represent up and down spin of an
electron. Methods for production and transport of spin entangled
electron in solid-state structures have been discussed in
Ref.~\cite{entangle}. 
For this input state electrons show
bunching behavior and the current shot noise in the interferometer
is enhanced \cite{bunching}. Electron bunching, in turn, leads to 
improvement in sensitivity. For each up or
down spin Fermi operators in this state scattering matrix
is given by Eq.~(\ref{smatrix}). The final state of the two electrons at
drains $\beta$ and $\delta$ is given by (for the qubit charge $i$)
\begin{eqnarray}
|\psi_f^i\rangle &=&\sqrt{2}\left[\bar{r}_i
\bar{t}^{\prime}_{i}c^{\dag}_{\beta\uparrow}
c^{\dag}_{\beta\downarrow}+\bar{t}_i
\bar{r}^{\prime}_{i}c^{\dag}_{\delta\uparrow}
c^{\dag}_{\delta\downarrow}\right. \label{fstate}\\ \nonumber
&&\left.+\frac{1}{2}(\bar{t}_i\bar{t}^{\prime}_i+\bar{r}_i\bar{r}^{\prime}_i)(c^{\dag}_{\beta\uparrow}
c^{\dag}_{\delta\downarrow}+ c^{\dag}_{\delta\uparrow}
c^{\dag}_{\beta\downarrow})\right]|F\rangle,
\end{eqnarray}
From this state one finds that the dephasing rate of the qubit
as~\cite{ynlee}
\begin{equation}
 \Gamma_d^s = \frac{eV}{h}\frac{(\Delta  T)^{2}}{T(1-T)} +
   4 \frac{eV}{h} T(1-T)(\Delta\zeta)^2
\label{eq:Gammads}
\end{equation}
The dephasing rate $\Gamma_d^s$ is enhanced by a factor of eight compared to
the case of injecting independent electrons at a single input 
(Eq.~(\ref{apdeph})).
Taking into account biasing two inputs with spin degeneracy in 
Eq.~(\ref{eq:Gammads}), the charge sensitivity (per electron) 
of the singlet state
is enhanced by a factor of two~\cite{note}.
The average current at the output $\beta$ or $\delta$ is independent
of the phase change $\Delta\phi$. In order to detect the phase shift 
$\Delta\phi$, it is necessary to measure shot noise or cross correlation
at the output leads.

In conclusion, we have discussed high sensitivity quantum limited
charge detection using electronic Fabry-Perot interferometer with
edge states. We note that in the realization of electronic
Mach-Zehnder interferometer significance of electron-electron
interactions at nonlinear bias~\cite{new} and temperature dependence
on dephasing~\cite{chung} have been reported. Such studies in our
scheme may also have experimental relevance.

We acknowledge helpful comments from A. Kolkiran.
This work was supported by 
the ``Cooperative Research Program" of
the Korea Research Institute of Standards and Science.


\begin{thebibliography}{99}
\bibitem{note} This is valid at low bias where the wave-packet size is larger than the distance between the two QPCs.
\end{thebibliography}
\end{document}